\documentclass[aps,pra,twocolumn,superscriptaddress,showpacs]{revtex4-1} 

\usepackage{amsmath}
\usepackage{graphicx}
\usepackage{amsfonts}
\usepackage{color}
\usepackage{amssymb}
\usepackage{mathrsfs}
\usepackage{multirow}

\newcommand{\avg}[1]{\left< #1\right>} 
\newcommand{\ket}[1]{\left|#1\right>} 
\newcommand{\bra}[1]{\left<#1\right|} 

\newcommand{\eref}[1]{Eq. (\ref{#1})}

\begin{document}

\title{Identical classical particles: half fermions and half bosons}\author{Falk T\"{o}ppel}
\email[]{falk.toeppel@mpl.mpg.de} 
\affiliation{Max Planck Institute for the Science of Light, G\"{u}nther-Scharowsky-Stra{\ss}e 1/Bldg. 24, 91058 Erlangen, Germany}
\affiliation{Institute for Optics, Information and Photonics, Universit\"{a}t Erlangen-N\"{u}rnberg, Staudtstra{\ss}e 7/B2, 91058 Erlangen, Germany}
\affiliation{Erlangen Graduate School in Advanced Optical Technologies (SAOT), Paul-Gordan-Stra{\ss}e 6, 91052 Erlangen, Germany}
\author{Andrea Aiello}
\affiliation{Max Planck Institute for the Science of Light, G\"{u}nther-Scharowsky-Stra{\ss}e 1/Bldg. 24, 91058 Erlangen, Germany}
\affiliation{Institute for Optics, Information and Photonics, Universit\"{a}t Erlangen-N\"{u}rnberg, Staudtstra{\ss}e 7/B2, 91058 Erlangen, Germany}
\date{\today}
\begin{abstract}
We study the problem of particle indistinguishability for the three cases known in nature: identical classical particles, identical bosons and identical fermions. By exploiting the fact that different types of particles are associated with Hilbert space vectors with different symmetries, we establish some relations between the expectation value of several different operators, as the particle number one and the interparticle correlation one, evaluated for states of a pair of identical (a) classical particles, (b) bosons and (c) fermions. We find that the quantum behavior of a pair of identical classical particles has exactly half fermionic and half bosonic characteristics.
\end{abstract}
%
\maketitle
\section{Introduction}

The either bosonic or fermionic character of particles available in nature fundamentally affects the way they behave when taking part in processes which have distinct outcomes, and each outcome may occur in several different manners. For example, in the celebrated Hong-Ou-Mandel experiment \cite{hong}, where two photons enter the input arms of a beam splitter (BS), the outcome when one has one photon in one output arm and one photon in the other arm, can occur in two different manners: Either both photons are reflected or are transmitted. The probability of such outcome is given by the modulus square of the \emph{sum} of the probability amplitudes of the two occurrences. Vice versa, when the same experiment is performed with electrons, which are fermions as opposed to photons that are bosons, the outcomes are completely different: The probability of having one electron in one output arm and one electron in the other arm, now is given by the modulus square of the \emph{difference} of the probability amplitudes of the two occurrences \cite{liu}. Also, the very quantum concept of entanglement is deeply influenced by the statistics of the particles \cite{banuls}. 
An overwhelmingly part of quantum information and quantum communication protocols \cite{nielsen}, quantum state tomography \cite{steinberg}, etc., make use of multiple-way processes, entangled states and classical communication. In this respect, a deeper understanding of the connection between bosonic, fermionic and classical nature of particles is highly desirable. 
In the present article we study this problem and we show that classical and quantum statistics of identical particles are not independent concepts but they have a common background. 

To begin with, we consider the article ``Fermion and boson beam-splitter statistics'' \cite{loudon}, in which Rodney Loudon examines the action of a BS on either a pair of bosons or a pair of fermions. The main results of this investigation are summarized in Tab. \ref{tab:result_moda}, reproduced below. It reports: (a) the expectation values for the number of particles $\langle\hat{n}^{(j)}\rangle$ at the two output ports $j\in\{1,2\}$ of the BS, (b) the variance $\langle[\Delta\hat{n}^{(j)}]^2\rangle=\langle[\hat{n}^{(j)}]^2\rangle-\langle\hat{n}^{(j)}\rangle^2$ of the output particle number and (c) the correlation $\langle\hat{n}^{(1)}\hat{n}^{(2)}\rangle$ between the particle numbers at the two output ports. These values are evaluated in three different cases: for a pair of identical (i) classical \cite{note1}, (ii) bosonic and (iii) fermionic ``particles'' \cite{note2}, entering the BS simultaneously from opposite sides. 
\begin{table}[h]
\caption{\label{tab:result_moda}Expectation values $\langle\hat{n}^{(j)}\rangle$, variances $\langle[\Delta\hat{n}^{(j)}]^2\rangle$ and correlation $\langle\hat{n}^{(1)}\hat{n}^{(2)}\rangle$ of particle numbers in the BS output arm $j\in\{1,2\}$, determined for two particles that enter the BS simultaneously from opposite sides.}
\begin{ruledtabular}
\begin{tabular}{@{}cccc}
   &  Classical  &  Bosonic  &  Fermionic  \\
\hline
$\langle\hat{n}^{(j)}\rangle$ & 1 & 1 & 1 \\
$\langle\bigl[\Delta\hat{n}^{(j)}\bigr]^2\rangle$ & $2|r|^2|t|^2$ & $4|r|^2|t|^2$ & 0 \\
$\langle\hat{n}^{(1)}\hat{n}^{(2)}\rangle$ & $|r|^4{+}|t|^4$ & $\bigl[|r|^2{-}|t|^2\bigr]^2$ & $\bigl[|r|^2{+}|t|^2\bigr]^2$ \\
\end{tabular}
\end{ruledtabular}
\end{table}
%
%

From Tab. \ref{tab:result_moda}, one can see that the result reported in the second column (referring to identical classical particles) can be obtained as the arithmetic mean of the corresponding results in the thurd and fourth (bosons and fermions). Synthetically one may say: identical classical particles = (bosons + fermions)/2. This result might appear surprising, based on the fact that particle indistinguishability is a very different concept in classical and quantum physics \cite{bach}. In this work we show that this remarkable feature has its origin in the special structure of the two-particle Hilbert space. Furthermore, we generalize our observation to multi-ports (MPs) \cite{jex,reck} and to a whole class of operators.


\section{Distinct two-particle states}
For the sake of simplicity, differently from Loudon, we do not treat the problem in the continuous infinite Hilbert space of quantum electrodynamics, but in a finite $d$-dimensional Hilbert space. Furthermore, as a generalization of the $2\!\times\!2$-port BS, we will examine an arbitrary unitary operation describing the action of a system with $d$ input modes and $d$ output modes upon a two-particle state ($d\!\times\!d$-port, or MP for short). 

We assume that a particle located in one particular mode $i\in\{1,...,d\}$ of the MP is described by the state vector $\ket{i}$. Moreover the $d$ modes are considered to be fully distinguishable, i.~e. $\avg{i|j}=\delta_{ij}$. Hence, these states form an orthonormal basis of a $d$-dimensional Hilbert space $\mathcal{H}=\mathrm{span}\{\ket{i}:i\in\{1,...,d\}\}$ describing all possible single-particle states. Here $\mathrm{span}\{ ... \}$ denotes the linear span of a set of vectors. 

In a next step we introduce states consisting of two identical particles. We define these distinct two-particle states $\ket{ij}=\ket{i}\otimes\ket{j}$ with $\ket{i},\ket{j}\in\mathcal{H}$ and $i,j\in\{1,...,d\}$, where $\otimes$ denotes the tensor or Kronecker product \cite{reed_simon_tensor}.
A few words about this notation that is used throughout the article are in order. Suppose, we have two distinct particles, say a red (R) one and a blue (B) one, each populating one mode (either input or output) of a $d$-dimensional system. So the state where the red particle is occupying mode $i$ while the blue particle is located in mode $j$ could be denoted indifferently either as $|R,i; B,j \rangle$ or as $|B,j ; R,i\rangle$. However, $|R,i; B,j \rangle$ and $|R,j ; B,i \rangle$ represent two physically distinct states because in the first case particle $R$ is in mode $i$ and particle $B$ in mode $j$, while in the second case modes have been swapped. This somewhat heavy notation can be greatly lightened by choosing a specific, although arbitrary, convention for the ordering of the two particles. Thus, in the following we use the convention that in $|i,j \rangle$ the first index (here $i$) always denotes the mode occupied by particle $R$, while the second index (here $j$) always refers to particle $B$, i.e., $|i,j \rangle$ and $|j,i \rangle$ correspond to $|R,i; B,j \rangle$ and $|R,j ; B,i \rangle$, respectively.

Distinct two-particle states are suitable to describe any pair of fully distinguishable particles, such as (i) two billiard balls, (ii) a horizontally and a vertically polarized photon, or (iii) two electrons with opposite spin, each partner being present in mode $i$ and $j$, respectively. Moreover, all distinct two-particle states are by definition orthonormal and thus do not interfere. In this sense they can be considered as describing classical particles. The distinct two-particle states form the standard basis of the $d^2$-dimensional Hilbert space describing all possible two-particle states: $\mathcal{H}^{(2)}=\mathcal{H}\otimes\mathcal{H}=\mathrm{span}\{\ket{ij}:i,j\in\{1,...,d\}\}$. Since $\mathcal{H}^{(2)}$ has finite dimension, all operators acting on it 
will be bounded operators \cite{reed_simon_bounded}.

Under the action of the MP, an input two-particle state undergoes a linear unitary transformation $\hat{U}^{(2)}:\mathcal{H}^{(2)}\rightarrow\mathcal{H}^{(2)}$, defined as $\hat{U}^{(2)}=\hat{U}\otimes\hat{U}$ with $\hat{U}:\mathcal{H}\rightarrow\mathcal{H}$ accounting for the linear unitary transformation of a single-particle state, i.e. $\hat{U}^{(2)}\ket{ij}=\hat{U}\ket{i}\otimes\hat{U}\ket{j}$ for all $i,j\in\{1,...,d\}$. 
The probability that the MP transforms a specific input state $\ket{\Phi}\in\mathcal{H}^{(2)}$ into a particular output state $\ket{\Psi}\in\mathcal{H}^{(2)}$ is, as usual, determined by $|\bra{\Psi}\hat{U}^{(2)}\ket{\Phi}|^2$.

\begin{figure}[h]
\centering
\includegraphics[width=3cm]{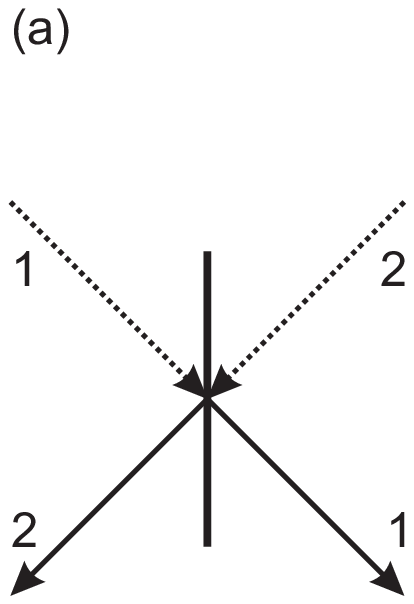}
\hspace{9mm}
\includegraphics[width=4.5cm]{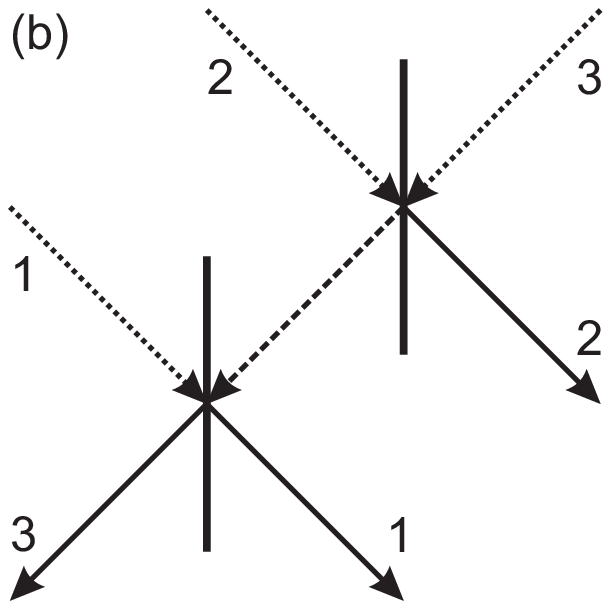}
\caption{Model systems built from BS with reflectivity $r$ and transmissivity $t$. Input and output modes are labeled by numbers. 
} \label{fig.examples} 
\end{figure}

Throughout this article we will exemplify our general considerations with the help of two MP models. The model systems are (a) a single BS ($d=2$) and (b) a combination of two BS ($d=3$), both depicted in Fig.~\ref{fig.examples}. 
By denoting with $|t|^2$ the probability that a particle in input mode $i$ is found in the same output mode and with $|r|^2$ the probability of ``flipping'', namely that a particle entering input mode $i$ is leaving output port $j$, one can find that the unitary operator $\hat{U}$ has the following matrix representation \cite{holbrow}
\begin{align*}
\begin{array}{c|c c }
  \hat{U} & \rotatebox{90}{$\ket{1}$} & \rotatebox{90}{$\ket{2}$}\\
  \hline
  \bra{1} & t & r \\
  \bra{2} & r & t \\
\end{array}\mathrm{~~in~case~(a)~and~~}
\begin{array}{c|c c c }
  \hat{U} & \rotatebox{90}{$\ket{1}$} & \rotatebox{90}{$\ket{2}$} & \rotatebox{90}{$\ket{3}$}\\
  \hline
  \bra{1} & t & r^2 & rt \\
  \bra{2} & 0 & t & r \\
  \bra{3} & r & rt & t^2 \\
\end{array}\mathrm{~~in~case~(b).}
\end{align*}
Here $r$ and $t$ are the complex reflectivity and transmissivity coefficients of the BS. They do obey the constraints
\begin{align}
\label{eq:bs_constraints}
|r|^2+|t|^2=1 \mathrm{~~and~~}rt^*+tr^*=0, 
\end{align}
ensuring $\hat{U}$ to be unitary and thus particle number conservation. Particularly in example (a) a Kronecker product yields the following matrix representation of 
$\hat{U}^{(2)}$:
\begin{align}
\label{eq:u2_distinct}
\begin{array}{c|c c c c}
  \hat{U}^{(2)} & \rotatebox{90}{$\ket{11}$} & \rotatebox{90}{$\ket{12}$} & \rotatebox{90}{$\ket{21}$} & \rotatebox{90}{$\ket{22}$} \\
  \hline
  \bra{11} & t^2 & rt & rt & r^2 \\
  \bra{12} & rt & t^2 & r^2 & rt \\
  \bra{21} & rt & r^2 & t^2 & rt \\
  \bra{22} & r^2 & rt & rt & t^2 \\
\end{array}.
\end{align}  

\section{Symmetric and antisymmetric subspace}
The distinct two-particle states introduced in the preceding section are not suitable for representing indistinguishable two-particle quantum states. According to quantum mechanics, such states are described by state vectors symmetric (bosons) or antisymmetric (fermions) under particle exchange. By using group theory, it is possible to show that the two-particle state space $\mathcal{H}^{(2)}$ can be expressed as direct sum of the symmetric subspace $\mathcal{H}_s\subset\mathcal{H}^{(2)}$ (indistinguishable bosons) and the antisymmetric subspace $\mathcal{H}_a\subset\mathcal{H}^{(2)}$ (indistinguishable fermions) \cite{herbut,landsman}, i.~e.~$\mathcal{H}^{(2)}=\mathcal{H}_s\oplus\mathcal{H}_a$. Following, we define these two subspaces as: 
\begin{subequations}
\label{eq:def_hs}
\begin{align}
\label{eq:def_hs_a}
\mathcal{H}_s&=\mathrm{span}\{\ket{s_{ij}}:i,j\in\{1,...,d\},j\geq i\}\\
\label{eq:def_hs_b}
&=\mathrm{span}\{\ket{s_{ij}}:i,j\in\{1,...,d\}\},
\end{align}
\end{subequations}
where $\ket{s_{ii}}=\ket{ii}$ and $\ket{s_{ij}}=(\ket{ij}+\ket{ji})/\sqrt{2}$ for $i\neq j$;
\begin{subequations}
\label{eq:def_ha}
\begin{align}
\label{eq:def_ha_a}
\mathcal{H}_a&=\mathrm{span}\{\ket{a_{ij}}:i,j\in\{1,...,d\},j>i\}\\
\label{eq:def_ha_b}
&=\mathrm{span}\{\ket{a_{ij}}:i,j\in\{1,...,d\},j\neq i\},
\end{align} 
\end{subequations}
with $\ket{a_{ij}}=(\ket{ij}-\ket{ji})/\sqrt{2}$ for $i\neq j$. Please note that the spanning sets given in \eref{eq:def_hs_a} and \eref{eq:def_ha_a} are bases of the subspaces, while the ones from \eref{eq:def_hs_b} and \eref{eq:def_ha_b} are overcomplete as $\ket{s_{ij}}=\ket{s_{ji}}$ and $\ket{a_{ij}}=-\ket{a_{ji}}$. However, we will often use the latter spanning sets since they simplify the notation a lot. 
Since $\mathcal{H}_s\perp \mathcal{H}_a$, joining the two bases from \eref{eq:def_hs_a} and \eref{eq:def_ha_a} yields a new orthonormal basis of $\mathcal{H}^{(2)}$. We will refer to it as the symmetric/antisymmetric basis.

We can define a linear operator $\hat{S}:\mathcal{H}^{(2)}\rightarrow\mathcal{H}_s$ projecting on the symmetric part and a linear operator $\hat{A}:\mathcal{H}^{(2)}\rightarrow\mathcal{H}_a$ projecting on the antisymmetric part of any two-particle state: $\hat{S}\ket{ij}=(\ket{ij}+\ket{ji})/2$ and $\hat{A}\ket{ij}=(\ket{ij}-\ket{ji})/2$ with $i,j\in\{1,...,d\}$. 
These two operators have the following important properties:
\begin{itemize}
\item The operators $\hat{S}$ and $\hat{A}$ are self-adjoint operators. 
\item The operators $\hat{S}$ and $\hat{A}$ are orthogonal ($\hat{S}\hat{A}=\hat{A}\hat{S}=0$) and hence do commute. 
\item The operators $\hat{S}$ and $\hat{A}$ are projectors 
and $\mathcal{H}_s$ ($\mathcal{H}_a$) is the eigenspace of $\hat{S}$ corresponding to the eigenvalue 1(0). The same property is obtained for $\hat{A}$ with the role of $\mathcal{H}_s$ and $\mathcal{H}_a$ swapped. 
\item The identity $\hat{S}+\hat{A}=1$ holds and the 
operator $\hat{S}-\hat{A}$ swaps the labeling of the two particles, i.e. 
\begin{align}
\label{eq:swapping_op}
(\hat{S}-\hat{A})\ket{ij}=\ket{ji}\mathrm{~~with~~} i,j\in\{1,...,d\}.
\end{align}
\end{itemize}

\section{Operators invariant under particle exchange}
Let us consider a linear operator $\hat{O}: \mathcal{H}^{(2)}\rightarrow\mathcal{H}^{(2)}$ that is invariant under particle exchange, namely an operator that satisfies the following equation: 
\begin{align}
\label{eq:op_inv}
\hat{O}=(\hat{S}-\hat{A})^\dagger\hat{O}(\hat{S}-\hat{A}),
\end{align}
where \eref{eq:swapping_op} should be remembered. Such operators treat the two identical particles forming a distinct two-particle state as indistinguishable. For example, although two photons identical except from orthogonal polarization are fully distinguishable, one cannot distinguish them without a polarizer.

Adding \eref{eq:op_inv} to the trivial identity $(\hat{S}+\hat{A})^\dagger\hat{O}(\hat{S}+\hat{A})=\hat{O}$, we derive
\begin{align}
\label{eq:SOS+AOA}
\hat{O}=\hat{S}^\dagger\hat{O}\hat{S}+\hat{A}^\dagger\hat{O}\hat{A}. 
\end{align}
Using this result and the properties 
of $\hat{S}$ and $\hat{A}$ mentioned above, we draw two conclusions: 
\begin{itemize}
\item The operator $\hat{O}$ maps a state from $\mathcal{H}_p$ always onto a state in $\mathcal{H}_p$ with parity $p\in\{s,a\}$. Consequently, $\hat{O}$ does not change the parity of symmetric and antisymmetric states.
\item The operator $\hat{O}$ does commute with $\hat{S}$ and $\hat{A}$.
\end{itemize}

Hence, we can introduce restrictions of $\hat{O}$ to the two subspaces $\mathcal{H}_s$ and $\mathcal{H}_a$. The operator $\hat{O}_s: \mathcal{H}_s\rightarrow\mathcal{H}_s$ shall be defined as $\hat{O}_s=\hat{S}^\dagger\hat{O}\hat{S}|_{\mathcal{H}_s}$ and $\hat{O}_a: \mathcal{H}_a\rightarrow\mathcal{H}_a$ via $\hat{O}_a=\hat{A}^\dagger\hat{O}\hat{A}|_{\mathcal{H}_a}$. Please note that the restrictions to the subspaces are necessary as the domain of $\hat{S}^\dagger\hat{O}\hat{S}$ and $\hat{A}^\dagger\hat{O}\hat{A}$ is whole $\mathcal{H}^{(2)}$. Thus, $\hat{O}_s$ and $\hat{O}_a$ are the equivalent of $\hat{O}$ in the bosonic and fermionic subspace, respectively. Additionally, the operator $\hat{O}$, with help of \eref{eq:SOS+AOA}, can  be put into the form
\begin{align}
\label{eq:op_decomp}
\hat{O}=\hat{S}^\dagger\hat{O}_s\hat{S}+\hat{A}^\dagger\hat{O}_a\hat{A}.
\end{align}
In summary any operator fulfilling \eref{eq:op_inv} consist of two parts acting solely either on the symmetric or antisymmetric subspace of $\mathcal{H}^{(2)}$.

An example of such kind of operator is the unitary transformation $\hat{U}^{(2)}=\hat{U}\otimes\hat{U}$, because remembering that $(\hat{S}-\hat{A})=(\hat{S}-\hat{A})^\dagger$ swaps the labeling of each two-particle state and $\hat{U}^{(2)}=\hat{U}\otimes\hat{U}$, we infer that:
\begin{align*} 
(\hat{S}-\hat{A})^\dagger\hat{U}^{(2)}(\hat{S}-\hat{A})\ket{ij}
&=(\hat{S}-\hat{A})^\dagger[\hat{U}\ket{j}\otimes\hat{U}\ket{i}]\\
&=\hat{U}\ket{i}\otimes\hat{U}\ket{j}=\hat{U}^{(2)}\ket{ij},
\end{align*}
for all $i,j\in\{1,...,d\}$. Therefore, due to the decomposition \eref{eq:op_decomp}, the matrix representation of $\hat{U}^{(2)}$ in the symmetric/antisymmetric basis contains, in contrast to \eref{eq:u2_distinct}, always two blocks of zeros, indicating forbidden transitions. The matrix representation of the restrictions $\hat{U}^{(2)}_s$ and $\hat{U}^{(2)}_a$ are the two non-zero blocks in the matrix representation of $\hat{U}^{(2)}$. In the case of our two model systems we find 
\begin{align*}
\begin{array}{c|c c c|c}
  \hat{U}^{(2)} & \rotatebox{90}{$\ket{s_{11}}$} & \rotatebox{90}{$\ket{s_{22}}$} & \multicolumn{2}{c}{\rotatebox{90}{$\ket{s_{12}}$} ~\qquad~ \rotatebox{90}{$\ket{a_{12}}$}} \\
  \hline
  \bra{s_{11}} & t^2 & r^2 & \sqrt{2}rt &  \\
  \bra{s_{22}} & r^2 & t^2 & \sqrt{2}rt & 0 \\
  \bra{s_{12}} & \sqrt{2}rt & \sqrt{2}rt & t^2+r^2 &  \\\cline{2-5}
  \bra{a_{12}} &   & 0 &   & t^2-r^2\\
\end{array}
\end{align*}
for model (a) and
\begin{align*}
\begin{array}{c|c c c c c c|c c c}
  \hat{U}^{(2)} & \rotatebox{90}{$\ket{s_{11}}$} & \rotatebox{90}{$\ket{s_{22}}$} & \rotatebox{90}{$\ket{s_{33}}$} & \rotatebox{90}{$\ket{s_{12}}$} & \rotatebox{90}{$\ket{s_{13}}$} & \multicolumn{2}{c}{\rotatebox{90}{$\ket{s_{23}}$} \quad~ \rotatebox{90}{$\ket{a_{12}}$}} & \rotatebox{90}{$\ket{a_{13}}$} & \rotatebox{90}{$\ket{a_{23}}$}\\
  \hline
  \bra{s_{11}} & t^2 & r^4 & r^2 t^2 & r C & tC & r^2 C &   &   &  \\
  \bra{s_{22}} & 0 & t^2 & r^2 & 0 & 0 & C &   &   &  \\
  \bra{s_{33}} & r^2 & r^2 t^2 & t^4 & r C & tC & t^2 C &   & \multirow{2}{*}{\quad0} &  \\
  \bra{s_{12}} & 0 & r C & r C & t^2 & rt & r C_+ &   &   &  \\
  \bra{s_{13}} & C & r^2 C & t^2 C & r C_+ & tC_+ & C^2 &   &   &  \\
  \bra{s_{23}} & 0 & tC & tC & rt & r^2 & tC_+ &   &   &  \\\cline{2-10}
  \bra{a_{12}} &   &   &   &   &   &   & t^2 & rt & {-}r C_- \\
  \bra{a_{13}} &   &   & \multicolumn{2}{c}{0}   &   &   & r C_- & tC_- & 0\\
  \bra{a_{23}} &   &   &   &   &   &   & {-}rt & {-}r^2 & tC_-\\
  \end{array}
\end{align*} 
for model (b), where we have defined $C=\sqrt{2}rt$ and $C_\pm=t^2\pm r^2$ for the sake of shortness.

By analyzing model system (a), one sees that two identical fermions entering the BS simultaneously in different modes, leave it in different modes: $|t^2-r^2|^2=1$ because of \eref{eq:bs_constraints}, thus obeying the Pauli exclusion principle \cite{liu}. For a 50/50 BS and two identical bosons in different input modes that arrive simultaneously at the BS, one obtains with help of \eref{eq:bs_constraints}: $|t^2+r^2|^2=0$. Therefore, they never exit the BS at different sides. This result is known as coalescence or Hong-Ou-Mandel effect \cite{hong}.

Considering example (b), the matrix element $\bra{a_{13}}\hat{U}^{(2)}\ket{a_{23}}$ vanishes since two identical fermions entering simultaneously mode 2 and 3 should take the same path when leaving the device in mode 1 and 3 but this is not permitted by Pauli's exclusion principle. For some transitions in model system (b) the Hong-Ou-Mandel effect can be observed as well, e.g., $|\bra{s_{12}}\hat{U}^{(2)}\ket{s_{13}}|^2=0$ for $|t|^2=|r|^2=1/2$.

As another example of an operator that does fulfill the identity \eref{eq:op_inv}, consider
\begin{align}
\label{eq:ph_number_modes}
\hat{n}^{(k)}&=\sum_{i=1}^d\bigl(\ket{ik}\!\bra{ik}+\ket{ki}\!\bra{ki}\bigr),
\end{align}
counting the number of particles in a particular mode $k\in\{1,...,d\}$ since 
$\hat{n}^{(k)}\ket{ij}
=(\delta_{kj}+\delta_{ki})\ket{ij}$.
Applying $\hat{n}^{(2)}$ for $d=2$, e.g., to the states $\ket{12}$ where mode 2 contains one particle and $\ket{22}$ with two particles in mode 2, yields $\hat{n}^{(2)}\ket{12}=\ket{12}$ and $\hat{n}^{(2)}\ket{22}=2\ket{22}$, as it should be. Note that all combinations of these number operators, like the correlation operator $\hat{n}^{(1)}\hat{n}^{(2)}$, are invariant under particle exchange as well. A further example is the operator 
\begin{align}
\label{eq:class_prob}
\hat{P}^{(kl)}=\frac{1}{1+\delta_{kl}}\bigl(\ket{kl}\!\bra{kl}+\ket{lk}\!\bra{lk}\bigr), 
\end{align}
whose expectation value with respect to a two-particle state determines the probability $P^{(kl)}=\langle \hat{P}^{(kl)}\rangle$ to find one particle in mode $k$ and the other in mode $l$. Consider for example the superposition $\ket{\psi}=(\ket{11}+\ket{12}+\ket{21})/\sqrt{3}$ for $d=2$. We attain $P^{(11)}
=1/3$, $P^{(12)}
=2/3$ and $P^{(22)}
=0$. 
The two observables 
$\hat{n}^{(k)}$ and $\hat{P}^{(kl)}$ obey a superselction rule since they do not connect the two subspaces $\mathcal{H}_s$ and $\mathcal{H}_a$ as we know from the general considerations above \cite{schweber}.

We obtain the following restrictions of $\hat{n}^{(k)}$ to the subspaces $\mathcal{H}_s$ and $\mathcal{H}_a$:
\begin{subequations}
\label{eq:ph_number_modes_quantum}
\begin{align}
\hat{n}^{(k)}_s&=\sum_{i=1}^d(1+\delta_{ik})\ket{s_{ik}}\!\bra{s_{ik}}\mathrm{~~and~~}\\
\hat{n}^{(k)}_a&=\sum_{i=1}^d(1-\delta_{ik})\ket{a_{ik}}\!\bra{a_{ik}},
\end{align}
\end{subequations}
as well as of $\hat{P}^{(kl)}$:
\begin{align}
\label{eq:quant_prob}
\hat{P}^{(kl)}_s=\ket{s_{kl}}\!\bra{s_{kl}}\mathrm{~~and~~}\hat{P}^{(kl)}_a=\delta_{kl}\ket{a_{kl}}\!\bra{a_{kl}}.
\end{align}

We now have all the ingredients at hand to explain the observation stated at the beginning of this article. Let us consider an observable $\hat{O}$ invariant under particle exchange, i.e. an operator that satisfies \eref{eq:op_inv}, and some distinct two-particle input state with both particles located in different modes $\ket{ij}$. 
Subjected to the MP, the input state is transformed into the output density operator $\hat{\rho}=\hat{U}^{(2)}{}^\dagger\ket{ij}\!\bra{ij}\hat{U}^{(2)}$. The expectation value of $\hat{O}$ with respect to this output state is $\langle\hat{O}\rangle=\mathrm{Tr}\bigl\{\hat{\rho}\hat{O}\bigr\}$. Recalling that the trace is defined as $\mathrm{Tr}\{...\}=\sum_{i=1}^d\bra{b_i}...\ket{b_i}$ for an arbitrary basis $\{\ket{b_i}:i\in\{1,...,d\}\}$ of $\mathcal{H}^{(2)}$, we obtain $\langle\hat{O}\rangle=\mathrm{Tr}_s\bigl\{\hat{\rho}\hat{O}\bigr\}+\mathrm{Tr}_a\bigl\{\hat{\rho}\hat{O}\bigr\}$ when using the symmetric/antisymmetric basis. Here we denoted tracing over the symmetric and antisymmetric basis as $\mathrm{Tr}_s\{...\}$ and $\mathrm{Tr}_a\{...\}$, respectively. In other words, we have exploited the fact that $\mathcal{H}^{(2)}=\mathcal{H}_s\oplus\mathcal{H}_a$ and $\mathcal{H}_s\perp\mathcal{H}_a$ to split the trace in $\mathcal{H}^{(2)}$ into two traces performed on the subspaces $\mathcal{H}_s$ and $\mathcal{H}_a$. Applying \eref{eq:op_decomp} yields
$\langle\hat{O}\rangle=\mathrm{Tr}_s\bigl\{\hat{\rho}\hat{S}^\dagger\hat{O}_s\hat{S}\bigr\}+\mathrm{Tr}_a\bigl\{\hat{\rho}\hat{A}^\dagger\hat{O}_a\hat{A}\bigr\}$, where the relation $\hat{S}\ket{\psi}=0$ for all $\ket{\psi}\in\mathcal{H}_a$ and the equivalent relation for $\hat{A}$ have been used. Expressing $\hat{U}^{(2)}$ by its restrictions to the subspaces $\mathcal{H}_s$ and $\mathcal{H}_a$ and by using some of the aforementioned properties of $\hat{S}$ and $\hat{A}$, we derive
\begin{align*}
\langle\hat{O}\rangle&=\mathrm{Tr}_s\bigl\{\hat{U}^{(2)}_s{}^\dagger[\hat{S}\ket{ij}\!\bra{ij}\hat{S}^\dagger]\hat{U}^{(2)}_s\hat{O}_s\bigr\}\nonumber\\
&\quad+\mathrm{Tr}_a\bigl\{\hat{U}^{(2)}_a{}^\dagger[\hat{A}\ket{ij}\!\bra{ij}\hat{A}^\dagger]\hat{U}^{(2)}_a\hat{O}_a\bigr\}. 
\end{align*}
With $\hat{S}\ket{ij}\!\bra{ij}\hat{S}^\dagger=\frac{1}{2}\ket{s_{ij}}\!\bra{s_{ij}}$ and $\hat{A}\ket{ij}\!\bra{ij}\hat{A}^\dagger=\frac{1}{2}\ket{a_{ij}}\!\bra{a_{ij}}$ we obtain the main result of this article:
\begin{align}
\label{eq:avg_relation}
\bra{ij}\!\hat{U}^{(2)}{}^\dagger\hat{O}\hat{U}^{(2)}\!\ket{ij}&=\frac{1}{2}\bra{s_{ij}}\!\hat{U}^{(2)}_s{}^\dagger\hat{O}_s\hat{U}^{(2)}_s \!\ket{s_{ij}}\nonumber\\
&\quad +\frac{1}{2}\bra{a_{ij}}\!\hat{U}^{(2)}_a{}^\dagger\hat{O}_a\hat{U}^{(2)}_a\!\ket{a_{ij}}.
\end{align}
By means of this, we have proven that after passing the MP, the expectation value of the observable $\hat{O}$, which is invariant under particle exchange, determined for a distinct two-particle state input (e.g., a pair of white billiard balls), with both partners located in different input modes, equals the arithmetic mean of the expectation values attained for the corresponding bosonic (e.g., a pair of identical photons) and fermionic (e.g., a pair of identical electrons) two-particle states with respect to the observables $\hat{O}_s$ and $\hat{O}_a$. 

If both input particles are located in the same mode, i.~e. $\ket{ii}$, we find immediately 
$\bra{ii}\hat{U}^{(2)}{}^\dagger\hat{O}\hat{U}^{(2)}\ket{ii}=\bra{s_{ii}}\hat{U}^{(2)}_s{}^\dagger\hat{O}_s\hat{U}^{(2)}_s\ket{s_{ii}}$.
In this case there is no difference in the behavior of identical bosons and identical classical particles. Note that due to Pauli's exclusion principle it is impossible to prepare two identical fermions simultaneously in the same mode.

\section{Discussion}
In Tab. \ref{tab:result_moda}, at the beginning of the article, some explicit values for the model systems (a) are presented. For their evaluation we considered input particles in mode 1 and 2. The observables are linear combinations of $\hat{n}^{(k)}$ (for classical particles) or $\hat{n}^{(k)}=\hat{n}_s^{(k)}$ and $\hat{n}^{(k)}=\hat{n}_a^{(k)}$ (for bosonic and fermionic particles), defined in \eref{eq:ph_number_modes} and \eref{eq:ph_number_modes_quantum}. Of course in this case one recovers the results obtained by R. Loudon \cite{loudon}.

Applying \eref{eq:avg_relation} to the operators $\hat{P}^{(kl)}$, $\hat{P}^{(kl)}_s$ and $\hat{P}^{(kl)}_a$ from \eref{eq:class_prob} and \eref{eq:quant_prob}, we obtain 
\begin{align*}
P^{(kl)}=\frac{1}{2}P^{(kl)}_s+\frac{1}{2}P^{(kl)}_a,
\end{align*}
when the input particles are located in two different modes $i$ and $j$. Thus, the transition probabilities at the MP between particular input and output modes for a pair of identical classical particles ($P^{(kl)}$) equals the arithmetic mean of the transition probability for bosons ($P^{(kl)}_s$) and fermions ($P^{(kl)}_a$). In this spirit, we conclude that a pair of identical classical particles ``acts'' half as a pair of bosons and half as a pair of fermions. In Tab. \ref{tab:result_modb} we give some probabilities for model system (b) to find one particle in mode $k$ and the other in mode $l$ when the input particles were prepared in modes $i$ and $j$. One verifies that averaging the last two columns (bosons and fermions), 
yields the second column (identical classical particles).

\begin{table}
\caption{\label{tab:result_modb}Probability to find a pair of particles in the output arms $k$ and $l$, when the two particles entered the model system (b) simultaneously in the input modes $i$ and $j$.}
\begin{ruledtabular}
\begin{tabular}{@{}cccc ccc}
   &   &   &   & Classical & Bosonic & Fermionic \\
 $i$ & $j$ & $k$ & $l$ & $\ket{ij}\rightarrow\ket{kl}$ or $\ket{lk}$ & $\ket{s_{ij}}\rightarrow\ket{s_{kl}}$ & $\ket{a_{ij}}\rightarrow\ket{a_{kl}}$ \\ 
 \hline
 1 & 2 & 2 & 3 & $|r|^2|t|^2$ & $|r|^2|t|^2$ & $|r|^2|t|^2$ \\
 1 & 3 & 1 & 3 & $|t|^2(|t|^4+|r|^4)$ & $|t|^2|t^2+r^2|^2$ & $|t|^2|t^2-r^2|^2$ \\
 2 & 3 & 1 & 3 & $2|r|^4|t|^4$ & $4|r|^4|t|^4$ & $0$ \\
 \end{tabular}
\end{ruledtabular}
\end{table}

Simulating fermionic behavior in a quantum walk with bosons can be realized by using entangled states of polarized photon pairs \cite{sansoni,matthews,spagnolo}. However, our results suggest in principle that, for example, from determining the probability distribution of two unitary quantum walks \cite{sansoni,matthews,spagnolo,silberhorn}, one performed with two distinguishable (e.g., orthogonally polarized) and the other with a pair of unentangled identical photons, the probability distribution of the same walk for a pair of indistinguishable fermions can be inferred. In this spirit one can simulate the behavior of unentangled fermions simply with unentangled bosons. Recently, boson sampling has been a fruitful area for experimental work \cite{}.

\section{Conclusion}
In summary, we have shown that the fermionic or bosonic nature of a pair of identical particles can be seen not only as a purely quantum feature, but it appears to be strictly connected to the statistical properties of classical particles. This result was achieved by means of a rigorous analysis of the well-established second-quantization formalism \cite{merzbacher} for a two-particle, $d$-mode system. For more than two particles the Hilbert space can no longer be decomposed as direct sum of a symmetric and an antisymmetric subspace. Nevertheless, generalizing some of these results to more than two particles is certainly possible, but some more advanced group theory tools, such as the Young tableaux, are required.


\begin{thebibliography}{35}
\bibitem{note1}
Think of, e. g., two white billiard balls entering a black box with two output ports. An observer who looks at the balls distribution at the two ports is not able to distinguish between the two balls since there is not any label attached to them: they are identical. However, the two particles are classical in the sense that if inside the black box occur different processes which yield to the same outcome A, then the probability of obtaining outcome A, is given by the sum of the probabilities of occurrence of each process leading to $A$. In other words, (identical) classical particles do not interfere.
\bibitem{note2}
Being aware of the particle-wave duality, we put the word particle in quotation marks. However, for the sake of brevity, we will omit the quotation marks in the remainder.
\end{thebibliography}
\end{document}